\documentclass[twocolumn,trackchanges,sort&compress,numbers]{aastex631}

\usepackage{graphicx}
\usepackage{epstopdf}
\usepackage{xcolor}
\usepackage{soul}

\usepackage{gensymb}

\begin{document}

\title{Updated kinematics of the Radcliffe Wave: non-synchronous, dipole-like vertical oscillations}

\author[0009-0001-7960-7512]{Zhi-Kai Zhu}
\affiliation{Guangxi Key Laboratory for Relativistic Astrophysics, School of Physical Science and Technology, Guangxi University, Nanning 530004, China}

\author[0000-0001-8060-1321]{Min Fang}
\affiliation{Purple Mountain Observatory, Chinese Academy of Sciences, 10 Yuanhua Road, Nanjing 210023, China}

\author[0000-0002-9138-5940]{Zu-Jia Lu}\thanks{luzujia@gxu.edu.cn}
\affiliation{Guangxi Key Laboratory for Relativistic Astrophysics, School of Physical Science and Technology, Guangxi University, Nanning 530004, China}

\author[0000-0001-6106-1171]{Junzhi Wang}
\affiliation{Guangxi Key Laboratory for Relativistic Astrophysics, School of Physical Science and Technology, Guangxi University, Nanning 530004, China}

\author[0000-0003-3144-1952]{Guang-Xing Li}\affiliation{South-Western Institute for Astronomy Research, Yunnan University, Kunming 650500, China}

\author[0009-0002-2379-4395]{Shiyu Zhang}
\affiliation{Purple Mountain Observatory, Chinese Academy of Sciences, 10 Yuanhua Road, Nanjing 210023, China}

\author[0000-0002-8898-1047]{Veli-Matti Pelkonen}
\affiliation{Institut de Ci\`{e}ncies del Cosmos, Universitat de Barcelona, IEEC-UB, Mart\'{i} i Franqu\`{e}s 1, E08028 Barcelona, Spain}

\author[0000-0002-5055-5800]{Paolo Padoan}
\affiliation{Institut de Ci\`{e}ncies del Cosmos, Universitat de Barcelona, IEEC-UB, Mart\'{i} i Franqu\`{e}s 1, E08028 Barcelona, Spain}
\affiliation{Department of Physics and Astronomy, Dartmouth College, 6127 Wilder Laboratory, Hanover, NH 03755, USA}

\author[0000-0002-7044-733X]{En-Wei Liang}
\affiliation{Guangxi Key Laboratory for Relativistic Astrophysics, School of Physical Science and Technology, Guangxi University, Nanning 530004, China}





\begin{abstract}

The kinematic information of the Radcliffe Wave (RW) is essential for determining its existence and gaining insights into its origin and evolution. In this work, we present an accurate measurement of the vertical velocity ($V_Z$) of RW by incorporating the radial velocity (RV) measures through two methods, which is crucial but was neglected previously. First, the velocities are measured towards young stars, using their RV measurements from APOGEE-2 and proper motion measurements from Gaia DR3. Second, we combine RV measurements toward clouds with proper motion measurements of associated Young Stellar Objects (YSOs) to determine the vertical velocities of the clouds. The results reveal that the oscillations in $V_Z$ are not synchronous with the vertical coordinate $Z$, which differs from the conclusions of previous studies. Instead, we find a 5 km$\cdot$s$^{-1}$$\cdot$kpc$^{-1}$ gradient in $V_Z$ along the RW, exhibiting a dipole-like pattern. Consequently, the kinematic arrangement does not show a corresponding coherence with the spatial arrangement, bringing the Radcliffe Wave model into question.

\end{abstract}


\section{Introduction} \label{Introduction}

The recent discovery of the Radcliffe Wave (RW) has sparked significant interest as it reveals a novel distribution pattern of molecular clouds and dense gas in the solar neighborhood. The RW was first proposed by \cite{Alves+2020Nature} as a coherent molecular cloud structure spanning 2.7 kiloparsecs in length, relying on the accurate distance measurements obtained by \cite{Zucker+2020A&A}, which belongs to a class of Glactic-scale gas filament reported since \cite{2013A&A...559A..34L}. It seems to exhibit some vertical oscillations in the Z direction of the Galactic coordinate system, which were suggested to be characterized as a damped sinusoidal mode.

As a so-called wave, the kinematics of RW is crucial for further understanding whether it exists as the real ``wave". However, it still remains uncertain. Recent studies by \cite{LiGuangXing+2022MNRAS} and \cite{Thulasidharan+2022A&A} have investigated the vertical oscillations of RW using tracers such as young stars and young clusters. \cite{LiGuangXing+2022MNRAS} found that the vertical velocity ($V_Z$) and vertical coordinate ($Z$) of RW oscillate synchronously, which means they exhibit a constant phase difference. The oscillations are consistent with the structure behaving either like a standing wave, or a wave traveling in the direction of decreasing $X^\prime$, which is defined in their work. While \cite{Thulasidharan+2022A&A} found a dipole in $V_Z$ along RW, which means the vertical velocities on the two sides of RW are comparable in magnitude but opposite in direction. However, in both works they made an approximation of ignoring the radial velocity (RV) when calculating $V_Z$, which is considerable and important for accurate kinematics. It will be considered in this work.

For precise measurements of the vertical oscillations of RW, we incorporate the RV component and calculate the $V_Z$ of the clouds using two approaches. In the first approach, we utilize multiple young star samples as tracers and obtain their RV measurements from the APOGEE-2 dataset \citep{Majewski+2017AJ}, as well as the proper motion and location measurements from the Gaia DR3 dataset \citep{GaiaDR3+2023A&A...674A...1G}. Combining all the above data, we derive the three-dimensional velocities of objects in Galactic coordinates. In the second approach, we directly obtain the RVs of clouds from surveys for CO and its isotope molecules \citep{Dame+2001ApJ,Zhang2024}, with the location information from the compendium provided by \cite{Zucker+2020A&A}. Subsequently, by incorporating the estimated proper motions of clouds based on associated young stars, we calculate the approximate velocities of molecular clouds within RW.

By utilizing these two methods, we aim to obtain more accurate measurements of the vertical oscillations of RW, leading to a comprehensive understanding of its kinematics and origin. The structure of the paper is as follows. In \S~\ref{data_method} we describe the data samples used in our analysis, following by our results in \S~\ref{results}. We discuss our main results in \S~\ref{discussion} and the main conclusions are summarized in \S~\ref{conclusion}.

\section{data and method} \label{data_method}

\subsection{Young Star Samples}
\label{YSOs_sample}

We use young star samples as tracers to study the kinematics of RW. First, we select a Young Stellar Object (YSO) sample from the catalogue provided by \cite{McBride+2021AJ}, combining it with the Class I/II YSO catalogue from \cite{Marton+2016MNRAS}. By cross-matching with Gaia DR3 within a 1 arcsec error, we obtain proper motion and parallax measurements for the YSOs. We then limit the YSOs to a rectangular region centered around RW, covering a vertical range from $Z$ = -300 pc to 250 pc in height, with a width of 500 pc and a length of 3.7 kpc. Subsequently, we exclude possible contamination in the YSO sample by referring to the method described in \cite{Zhangmm+2023ApJS}. Finally, after removing 1643 repeated YSOs, we obtain a YSO sample containing 103680 members.

Additionally, we incorporate a young open cluster sample (younger than 100 Myr) from the open cluster catalogue by \cite{Cantat-Gaudin+2020A&A} and an OB star sample from the Alma catalogue of OB stars by \cite{Pantaleoni+2021MNRAS}. These two samples are limited to the same rectangular region and updated with  DR3 information. The young open cluster sample contains 10735 cluster members, 3406 of which overlap with the YSO sample. The OB star sample contains 1115 members, with 55 overlapping with previous samples.

\subsection{Radial Velocity}
\label{radial_velocity}

In the first method, young stars are utilized as tracers. To derive RV of young star samples, we perform a cross-match within 1 arcsec tolerance between the young stars and the Sloan Digital Sky Survey IV (\textit{SDSS-IV}; \citealp{Blanton+2017AJ}) APOGEE-2 \citep{Majewski+2017AJ} dataset. 
We utilize RV measurements from APOGEE instead of Gaia, as its accuracy for YSOs is higher than Gaia's \citep{Kounkel+2023ApJS}. The comparison of the two datasets for the YSO sample are shown in Appendix \ref{app_b}. However, the coverage of RV measurements for YSOs is quite limited.

\begin{figure}[t]
\centering
\includegraphics[width=1\linewidth]{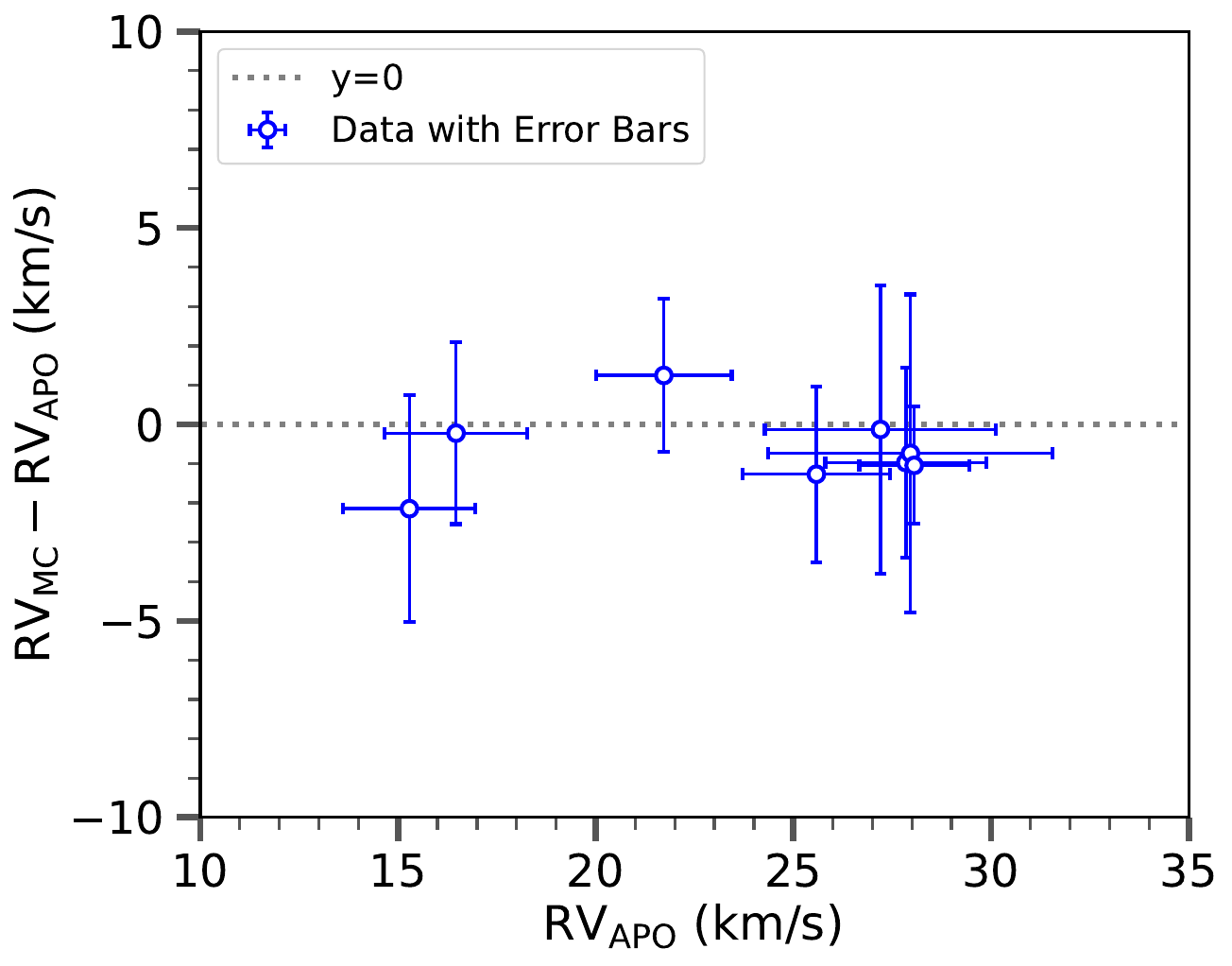}
\caption{A comparison of RV between clouds and their associated YSOs. Each data point represents a cloud, where the x-axis represents the median RV obtained from APOGEE for its associated YSOs (${\mathrm{RV}}_{\mathrm{APO}}$), and the y-axis represents the residual between the average RV of clouds (${\mathrm{RV}}_{\mathrm{MC}}$) and ${\mathrm{RV}}_{\mathrm{APO}}$. Only clouds with more than 10 associated YSOs having RV measurements are shown in this figure.}
\label{Figure_5}
\end{figure} 

\begin{figure*}[htbp]
\centering
\includegraphics [width=0.9\linewidth]{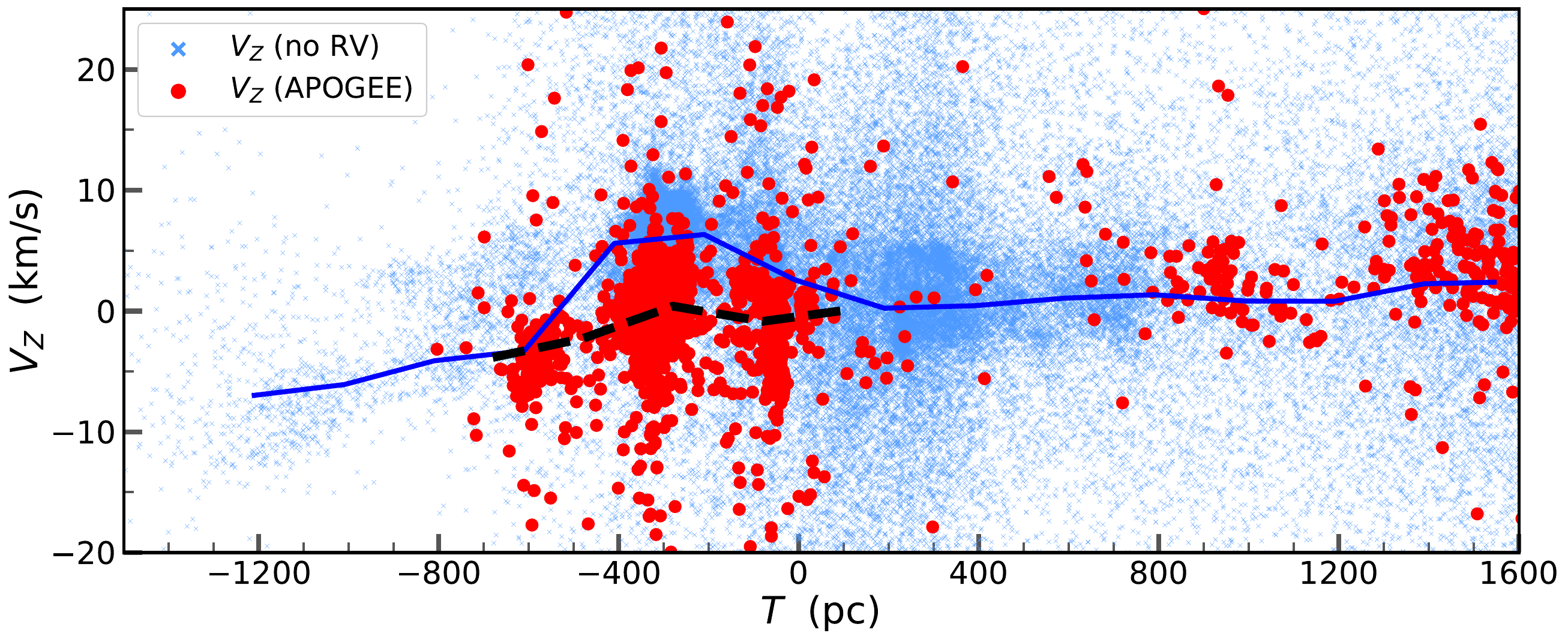}
\caption{$V_Z$ distribution of YSOs along the Radcliffe Wave. The red circles represent more accurate $V_Z$ corrected by RV from APOGEE, while the blue crosses indicate approximate $V_Z$ values calculated without considering RV.} $T$ represents the direction that the Radcliffe Wave extends. The blue lines represent the median values of blue crosses within 200 pc bins along the $T$ axis. The black dashed lines represent the median values of the red circles. We only include bins that contain more than 30 stars to ensure statistical significance in our analysis. 
\label{Figure_2}
\end{figure*}

\begin{figure*}[htbp]     
\centering
\includegraphics[width=1\linewidth]{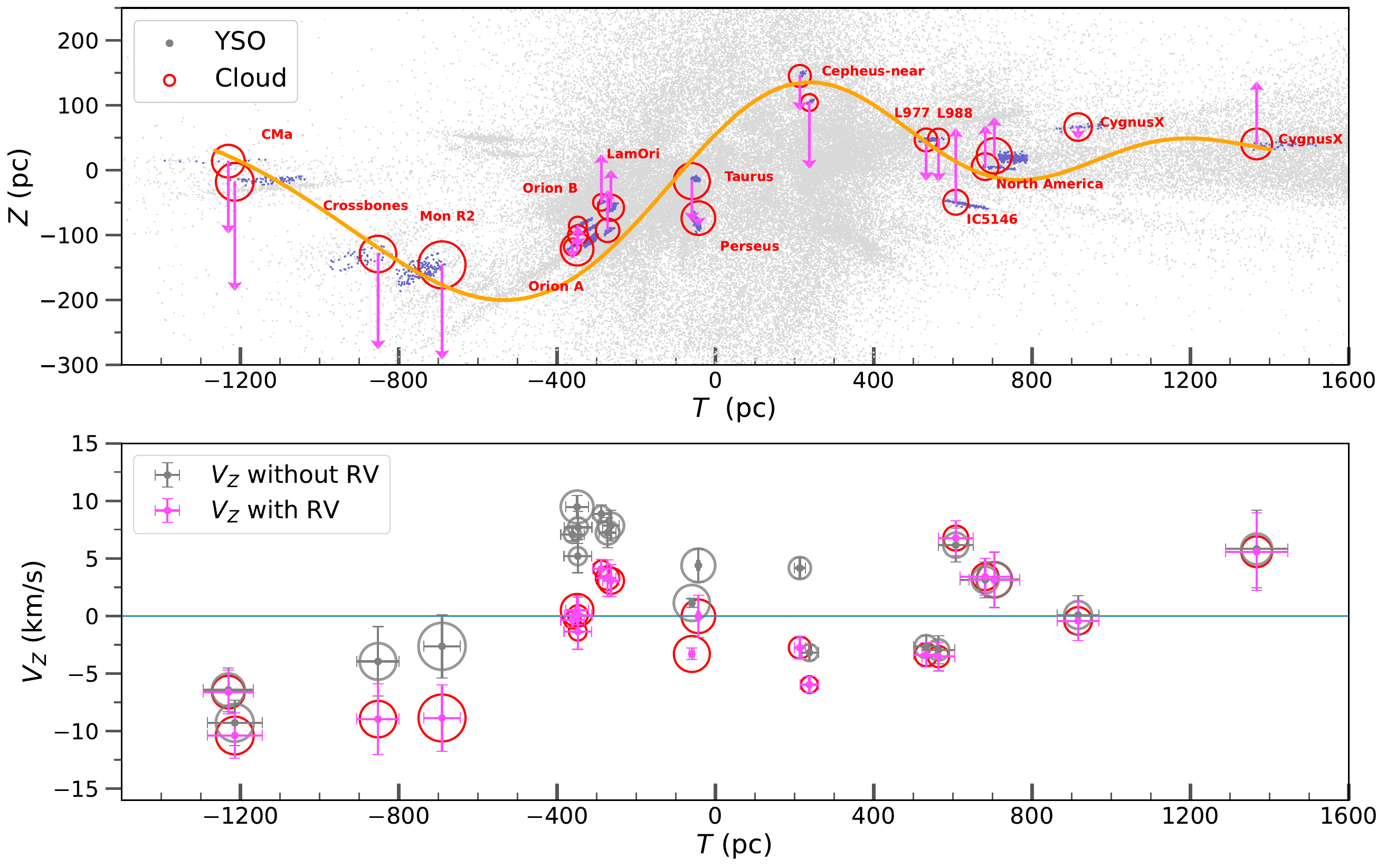}
\caption{The spatial and $V_Z$ distributions of the cross-matched molecular clouds along the Radcliffe Wave. The matched parts of clouds are depicted as red circles, where the radii represent the relative sizes of the matched parts based on the $T$ axis. In the upper panel, $Z$ is the vertical coordinate in the Galactic coordinate system, while the red circles indicate the cloud locations and the pink arrows indicate the directions and magnitudes of $V_Z$. The length of each arrow corresponds to the value of $V_Z$ (in km/s) multiplied by 15. The orange curve marks the fit to the position of the Radcliffe Wave model suggested by \cite{Alves+2020Nature}. In contrast, locations of the YSOs in the YSO sample are represented by grey points, and the purple ones indicate the cross-matched YSOs. The lower panel displays the distribution of $V_Z$. The grey circles and red circles represent the $V_Z$ results obtained without and with RV information, respectively. Error bars are included to indicate the 1$\sigma$ dispersion of $V_Z$ and the uncertainties of $T$, which incorporate both statistical uncertainties and systematic uncertainties.}
\label{Figure_3}
\end{figure*}

\begin{figure}
\centering
\includegraphics[width=\linewidth]{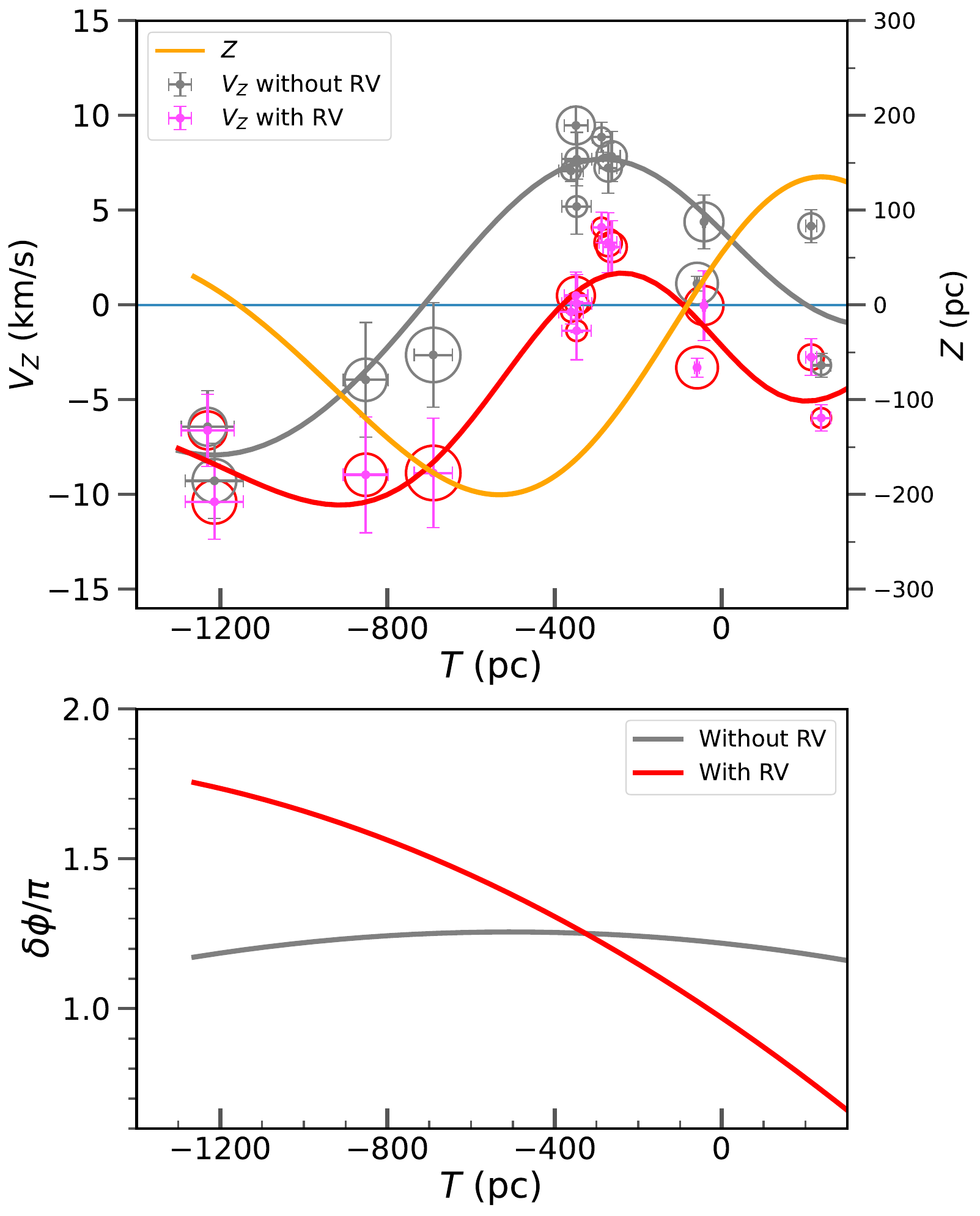}
\caption{Fitting results of the spatial and $V_Z$ distribution of clouds in the $T$<300~pc region. In the upper panel, the grey curve and red curve are the fitting results of $V_Z$ without and with RV measurements, respectively. The orange curve is the fitting result of  $Z$ (right $y$-axis). The lower panel shows the phase difference ($\delta\phi$) between $Z$ and $V_Z$, with the grey and red curves representing the distribution of $\delta\phi$ without and with RV measurements, respectively.}
\label{Figure_4}
\end{figure}

\begin{figure*}
\centering
\includegraphics[width=1\linewidth]{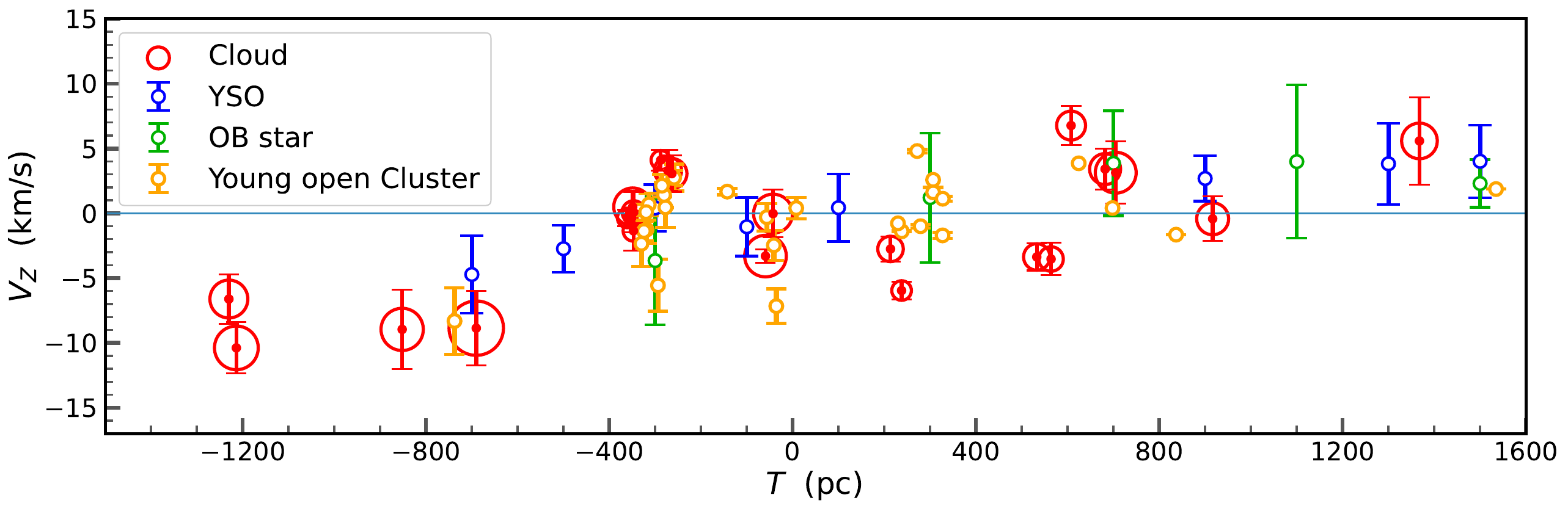}
\caption{The $V_Z$ distribution of clouds (calculated using RV from CO surveys) and young star samples (calculated using RV from APOGEE), with error bars indicating  1$\sigma$ dispersions.}
\label{Figure_6}
\end{figure*}

In the second method, the aim is to obtain RV measurements of the molecular clouds within a larger coverage directly. We use RV measurements from the whole-Galaxy CO surveys by \cite{Dame+2001ApJ}, which were observed by using the 1.2-meter Millimeter-Wave Telescope at the CfA and its counterpart on Cerro Tololo in Chile. These surveys cover the majority of molecular clouds within RW. 
The CO surveys provide RV measurements but lack the information of proper motion of the clouds, which is still a challenge to observe directly. Fortunately, there are strong kinematic correlations between the young stellar populations and associated molecular gas and clouds, as reported in \cite{Furesz+2008ApJ}, \cite{Tobin+2015AJ} and \cite{FangMin+2017AJ}. These studies provided strong evidences to support that young stars inherit RV from the molecular clouds, implying there is also a similarity in their proper motions. In this approach, we acquire approximate proper motions for certain molecular clouds from associated YSOs in the YSO sample. This approximation allows us to calculate $V_Z$ for the majority of clouds within RW. To obtain more precise measurements of RV and distance for the Cygnus X and North America regions, we also rely on the data (\citealp{Zhang2024}) from the Milky Way Imaging Scroll Painting (MWISP) project, which is a multi-line survey in $^{12}$CO/$^{13}$CO/C$^{18}$O along the northern Galactic plane with PMO-13.7m telescope \citep{Suyang2019}. More details of the method are described in Appendix \ref{app_a}.

To estimate the discrepancy of the RV between the clouds and their associated YSOs, we calculate the median RV obtained from APOGEE for its associated YSOs (${\mathrm{RV}}_{\mathrm{APO}}$), and the residual between the average RV of clouds (${\mathrm{RV}}_{\mathrm{MC}}$) and ${\mathrm{RV}}_{\mathrm{APO}}$, as shown in Figure \ref{Figure_5}. We found that there is a strong agreement between the two velocities, as evidenced by a maximum absolute residual of only 2.14 km/s and a high correlation coefficient of 0.98. This strong correlation suggests that the second method illustrated above is indeed feasible.

By combining RV with the proper motion and location data, along with the known component of Solar motion $(U_\odot,V_\odot,W_\odot) = (11.1, 12.24, 7.25) \,\rm km \,\rm s^{-1}$ \citep{Ralph_Schönrich+LSR+2010MNRAS}, we are able to derive the three-dimensional velocities of objects. Additionally, we adopt the precise measurement of the Sun's position, $Z_\odot=20.8 \,\rm pc$, from \cite{Bennett+Z_sun+2019MN} to obtain the vertical coordinate (Z) in the Galactic coordinate system.

\section{results}
\label{results}

Initially, we utilize RV measurements from APOGEE in the first method and obtain more accurate vertical velocities of the YSO sample. These updated $V_Z$ values are shown with the red circles in Figure \ref{Figure_2}. For comparison, we use blue crosses to represent the approximate $V_Z$ values calculated without RV measurements. In this work, the Galactic coordinate system is rotated $30\degree$ anticlockwise with respect to the $Z$ axis, aligning the rotated $Y$ axis (hereafter $T$) with the direction of RW extension. In Figure \ref{Figure_2}, we find that considering RV generally leads to smaller $V_Z$ values. This attribute to the RV component in the $Z$ direction (${\mathrm{RV}}\cdot\sin b$; $b$ is the Galactic latitude). As illustrated in Figure \ref{Figure_2}, this component is significant and comparable to the amplitude of $V_Z$, particularly in the solar neighborhood (around $T=0$) where stars typically have higher Galactic latitudes. As a result, the peak of the approximate oscillation (-500~pc<$T$<0~pc, corresponding to 0.75~kpc<$X^\prime$<1.25~kpc in Figure 3 in \citealp{LiGuangXing+2022MNRAS}) tends to flatten when RV is taken into account. This suggests that the synchronous oscillation reported in previous studies may arise from the neglect of RV, and incorporating RV measurements is necessary. However, the coverage of RV measurements for YSOs is primarily limited to nearby regions.

To gain a larger coverage and more comprehensive understanding, we directly study the molecular clouds as a complementary approach. In this approach, we combine the RVs of clouds and the estimated proper motions and locations based on associated YSOs in the YSO sample. As a result, we obtain the spatial distribution and $V_Z$ distribution of the crossed-matched clouds, as the red circles illustrated in Figure \ref{Figure_3}. As shown in the lower panel, the $V_Z$ values vary between -10.4~km/s and 6.8~km/s, with a dispersion of about 2~km/s. Compared to the results without RV indicated by grey circles, incorporating the RV components leads to smaller values of $V_Z$ in the $T$<300~pc region, while there is little change in other regions. Thus, the kinematic pattern of RW is different and should be reconsidered. Besides, the revised $V_Z$ demonstrates an overall upward trend as $T$ increases. 

For investigating the potential synchronous oscillations between $V_Z$ and $Z$, we perform fitting analyses on $V_Z$ and $Z$ distribution from Figure \ref{Figure_3}. We adopt the function proposed by \cite{Alves+2020Nature} in their equation (1) for our fitting, as it effectively explained the distribution of $Z$. We focus on the $T$<300~pc region, where \cite{LiGuangXing+2022MNRAS} suggested the presence of synchronous oscillations, which requires a constant phase difference. Figure \ref{Figure_4} displays the fitting results. 
In the upper panel, the grey curve obtained without RV resembles the yellow line shown in Figure 3 of \cite{LiGuangXing+2022MNRAS}. However, when the RV components are taken into account, as indicated by the red fitting curve, the values of $V_Z$ generally decrease in this region. Although a wave peak still exists in the -500~pc<$T$<0~pc region, inheriting a similar $T$ coordinate, the wave trough around $T=-1200$~pc moves to around $T=-900$~pc. This leads to a shorter period and a faster-changing phase of the red curve compared to the grey one. As a result, as shown in the lower panel, there is a change in $\delta\phi$, the phase difference between $Z$ (orange curve in the upper panel) and $V_Z$. When calculated without RV measurements, $\delta\phi$ appears constant; however, it decreases as $T$ increases when RV is taken into account. These results indicate that the previously reported synchronous oscillations arise from the neglect of the RV component, and there are no synchronous oscillations. In addition, separate K-S tests for the two $V_Z$ data groups yielded p-values of 0.94 for both, which indicates excellent fitting performance.


Furthermore, we study the $V_Z$ distribution of the OB star sample and the young open cluster sample. We analyze the median $V_Z$ for each cluster, as well as the median $V_Z$ of YSOs and OB stars in bins of 200 pc along the $T$ axis, keeping the bins containing more than 10 stars.
As a result, Figure \ref{Figure_6} depicts the $V_Z$ distribution of clouds and young star samples. 
In Figure \ref{Figure_6}, despite some minor oscillations in the intermediate range, there is an overall upward trend in $V_Z$ with increasing $T$. The trend is consistent among different samples, and $V_Z$ values exhibit an average gradient around 5 km$\cdot$s$^{-1}$$\cdot$kpc$^{-1}$ in the $T$ direction. The distribution of $V_Z$ resembles a dipole centered at $T=0$.

\section{discussion} \label{discussion}

\begin{figure*}[t]
\centering
  \includegraphics[width=0.9\linewidth]{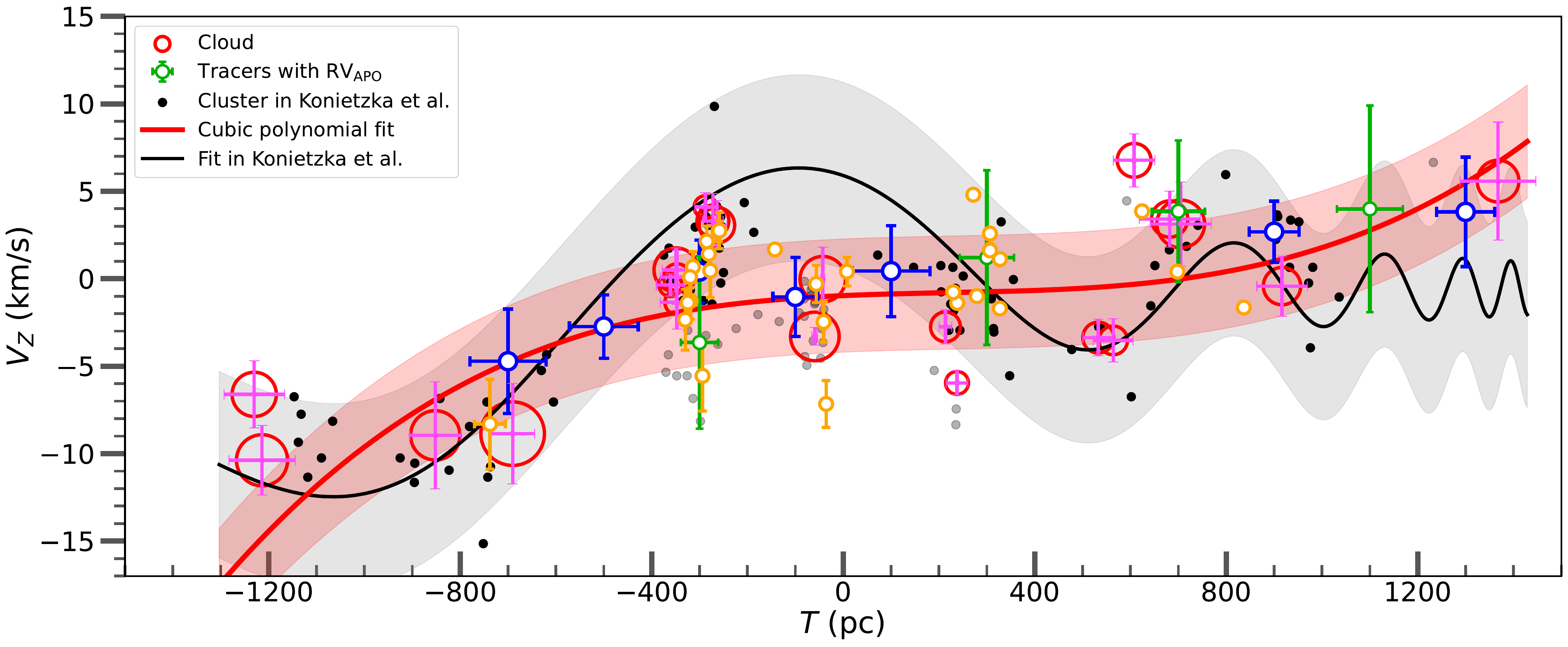}\label{Figure7a} \\ 
  \includegraphics[width=0.48\linewidth]{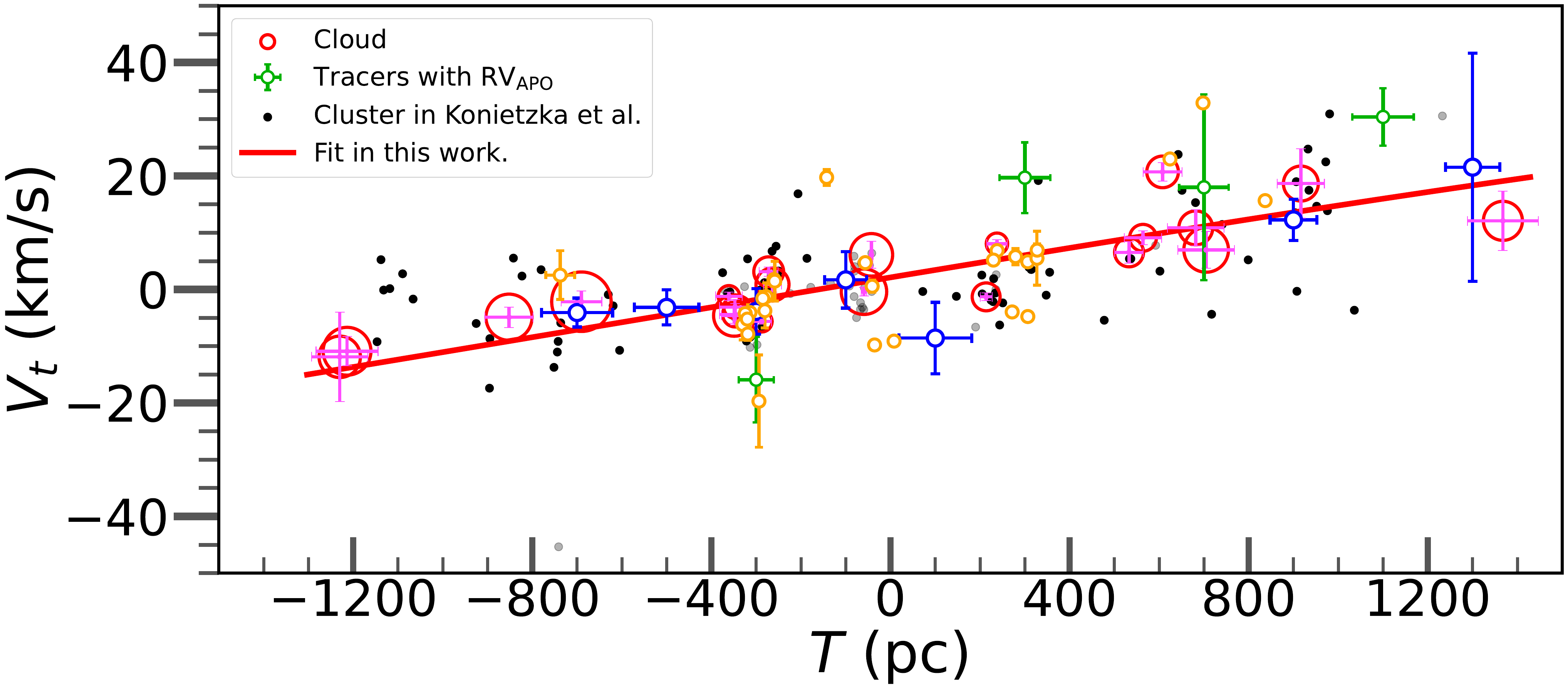}\label{Figure7b} 
  \hspace{0.5cm}
  \includegraphics[width=0.48\linewidth]{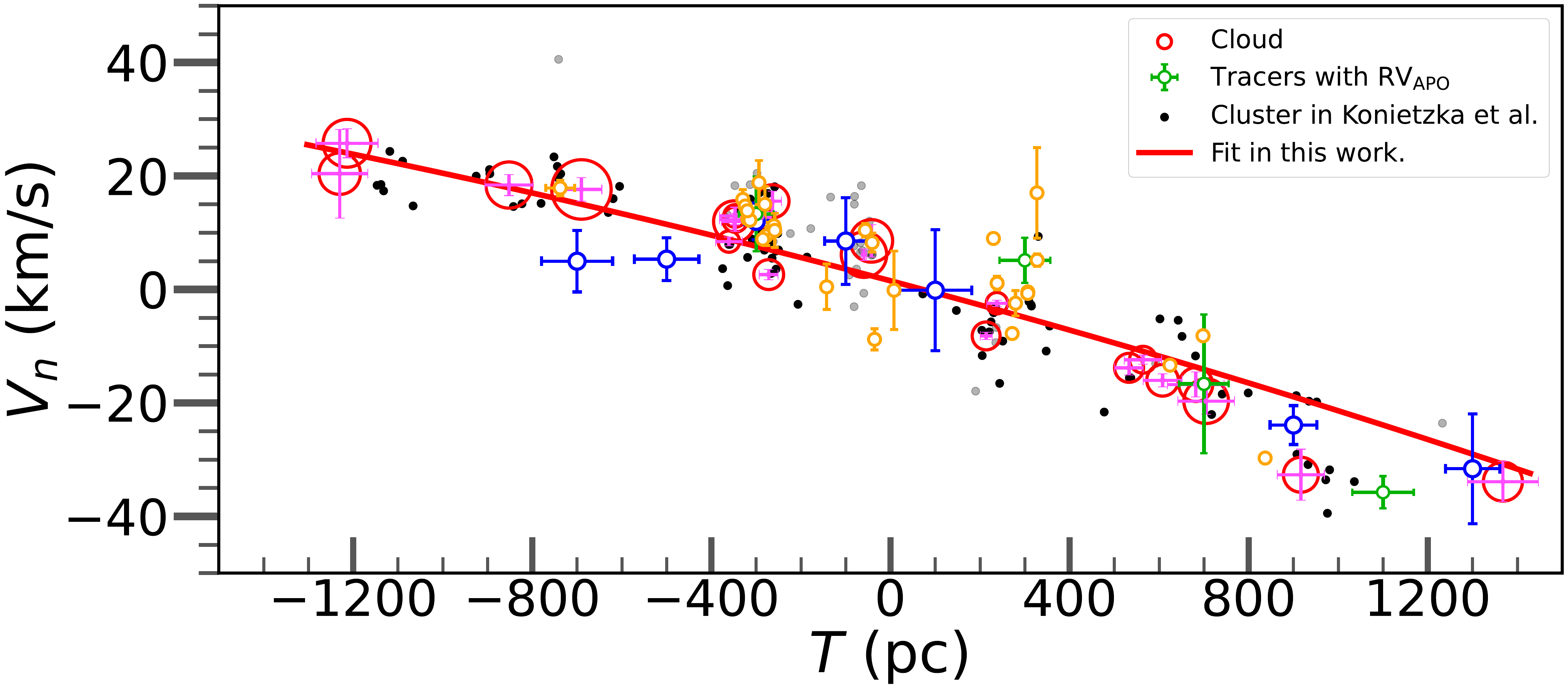}\label{Figure7c} 

\caption{Velocity distribution in the Local Standard Rest (LSR) frame versus the results in \cite{Konietzka+2024Nature} and their fitting model. Red circles represent the results of clouds, along with other tracers using RV from APOGEE in this study. These tracers are indicated by blue, green and yellow points, representing YSOs, OB stars and young open clusters, respectively. For comparison, the results of clusters from \cite{Konietzka+2024Nature} are depicted as black points, with grey points indicating outliers as defined in their research. The upper panel: the $V_Z$ distribution. The red line represents the fitting result using cubic polynomial in this study. The shaded areas indicate the 1$\sigma$ ranges of errors for both fitting models. The lower two panels: the distributions of $V_t$ and $V_n$ with fitting lines which are consistent with models in \cite{Konietzka+2024Nature}, respectively.}
\label{Figure_7}
\end{figure*}

After correcting $V_Z$ for the RW using two methods, we obtain a more accurate and comprehensive perspective of the $V_Z$ distribution. Different from the previous conclusion, we believe that the vertical oscillation in RW is non-synchronous. The discrepancy can be attributed to the RV component. However, we also note that different authors have adopted distinct models, which may lead to a different conclusion. Recently, \cite{Konietzka+2024Nature} proposed that the structure of RW has synchronous oscillations. Their conclusions were based on fitting another sinusoidal function to the $V_Z$ distribution, with certain regions treated as outliers, such as Taurus, Perseus, IC5146, and some parts of Orion, Cepheus-near and Cygnus X, which are included in the fitting procedure in our work. 
We compare our results with theirs in Figure \ref{Figure_7}. In the upper panel, we attempt to fit the $V_Z$ distribution using a cubic polynomial and find that this model exhibits a smaller fitting error and requires much fewer parameters. In the lower two panels, we also investigate velocities parallel to the Galactic disk and divide them into two components: the velocity in the $T$ direction ($V_t$) and the velocity perpendicular to both the $T$ and $Z$ directions, away from the Galactic center ($V_n$). These results are consistent with those from \cite{Konietzka+2024Nature}. Hence, we adopt their fitting approach in these two directions. Despite their explanation that the defined outliers in $V_Z$ could be attributed to feedback from expanding bubbles, no corresponding velocity deviation was observed in the other two directions.

Additionally, the revised $V_Z$ exhibits an upward trend as $T$ increases, with an average $V_Z$ gradient around 5 km$\cdot$s$^{-1}$$\cdot$kpc$^{-1}$ in the $T$ direction. This trend reveals a dipole pattern centered at $T=0$ in the distribution of $V_Z$, similar to the finding in \cite{Thulasidharan+2022A&A}, but with a larger velocity gradient. The discrepancy can be attributed to our improvement by considering RV measurements, which result in lower $V_Z$ values within the $T<300$ pc range, but comparable $V_Z$ values in the $T>300$ pc range. Besides, the dipole pattern is consistent among different star samples, implying the origin of this pattern has comparable effects on these samples. This kinematic pattern of RW is not consistent with its wave-like spatial pattern in the $Z$ direction, casting doubt on the RW model and even its actual existence.

\section{conclusion}\label{conclusion}

We analyze the kinematics of the Radcliffe Wave, incorporating radial velocity measurements through two methods. The radial velocity is significant for its kinematics but has been ignored previously. We find that the oscillations of $V_Z$ in the Radcliffe Wave are not synchronous with its vertical coordinate $Z$. Instead its $V_Z$ distribution shows a dipole-like pattern with a 5 km$\cdot$s$^{-1}$$\cdot$kpc$^{-1}$ $V_Z$ gradient in the direction of its extension. Consequently, the Radcliffe Wave's kinematics does not show a corresponding coherence with its wave-like spatial arrangement, bringing its model into question. 

\section{Acknowledgements}

This research made use of the data from the Milky Way Imaging Scroll Painting (MWISP) project, which is a multi-line survey in $^{12}$CO/$^{13}$CO/C$^{18}$O along the northern galactic plane with PMO-13.7m telescope. We are grateful to all the members of the MWISP working group, particularly the staff members at PMO-13.7m telescope, for their long-term support. MWISP was sponsored by National Key R$\&$D Program of China with grants 2023YFA1608000 $\&$ 2017YFA0402701 and by CAS Key Research Program of Frontier Sciences with grant QYZDJ-SSW-SLH047. ZJL acknowledges support by the grant AD23026127 and 2024GXNSFBA010436, funded by Guangxi Science and Technology Project. GXL acknowledges support from NSFC grant No. 12273032 and 12033005. VMP and PP acknowledge financial support by the grant PID2020-115892GB-I00, funded by MCIN/AEI/10.13039/501100011033. This work is supported by the National Natural Science Foundation of China (Grant No.12133003). This work is also supported by the Guangxi Talent Program (“Highland of Innovation Talents”).

\bibliography{reference}{}

\bibliographystyle{aasjournal}

\newpage 

\appendix

\section{Cross match with CO surveys}
\label{app_a}

\setcounter{figure}{0} 
\renewcommand\thefigure{A\arabic{figure}} 

\begin{figure}[htbp]
\centering
\includegraphics[width=0.79\linewidth]{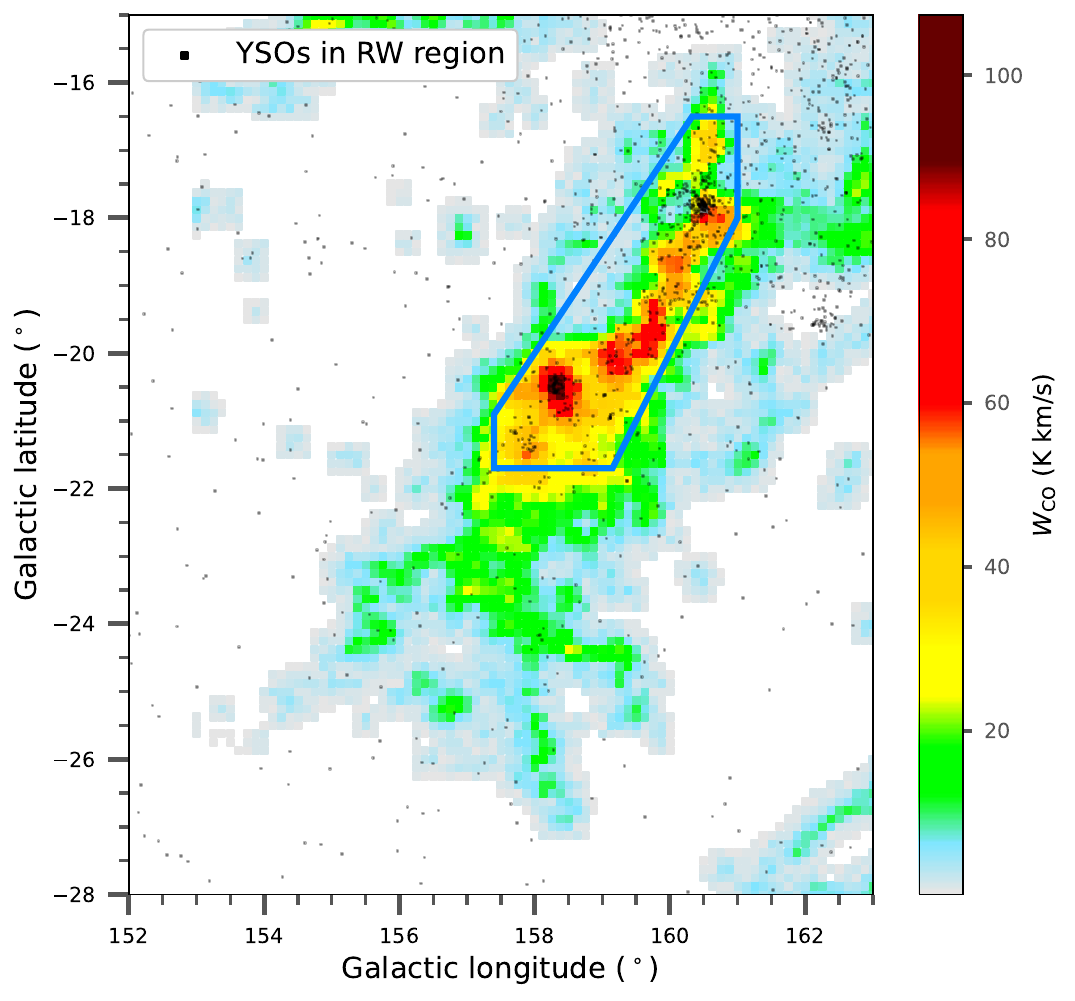}
\caption{Velocity-integrated spatial CO Map of the Perseus area. YSOs are indicated by black points, marking their respective coordinates within the map. The relatively associated region is marked by the enclosing blue folding frame.}
\label{appendix_A}
\end{figure}

When combining the YSO sample with RV measurements of molecular clouds from the whole-Galaxy CO surveys conducted by \cite{Dame+2001ApJ}, our focus is on the regions where YSOs and clouds with higher density (corresponding to larger velocity-integrated intensity of CO—$W_{\mathrm{CO}}$) are associated. 

For the spatial cross-matching process,  we initially match associated areas in the velocity-integrated spatial ($l$-$b$) CO map based on Galactic latitude ($b$) and Galactic longitude ($l$). This enables us to identify relatively associated regions, such as the area enclosed by the blue frame depicted in Figure \ref{appendix_A} for the Perseus molecular cloud region. 

To obtain more accurate RV measurements, we eliminate redundant peaks that may be attributed to overlapping clouds with similar latitude and longitude but at different distances, with reference to previous works, such as \cite{Dame+Persues+2023ApJ} for the Perseus molecular cloud area. For other clouds, we also refer to several articles including \cite{Maddalena+Orion(1)+1986ApJ,Leung+CygnusX(1)+1992ApJS,Dobashi+L988+1994ApJS,Lang+LamOri+2000A&A,Wilson+Orion(2)+2005A&A,Xu+L977L988(1)+2013ApJ,Da+Rio+OrionA(1)+2017ApJ,Duan+Taurus+2023RAA}.

Additionally, we determine the distances of the clouds by referring to the compendium provided by \cite{Zucker+2020A&A}. For more accurate RV and distance measurements of Cygnus X and the North America region, we also refer to the data (\citealp{Zhang2024}) from the MWISP project. Thereafter, we utilize the parallax measurements of YSOs obtained from Gaia DR3. This allows us to refine the distance estimation and effectively eliminate YSOs that are not spatially correlated with the molecular clouds in the radial direction.

Subsequently, we combine RV measurements of the associated cloud areas with the median value of the proper motion measurements of YSOs within the same areas. This combination enables us to calculate the vertical velocities of the clouds.

\section{APOGEE vs Gaia}
\label{app_b}

\setcounter{figure}{0} 
\renewcommand\thefigure{B\arabic{figure}} 

\begin{figure}[htbp]
\centering
\includegraphics[width=0.68\linewidth]{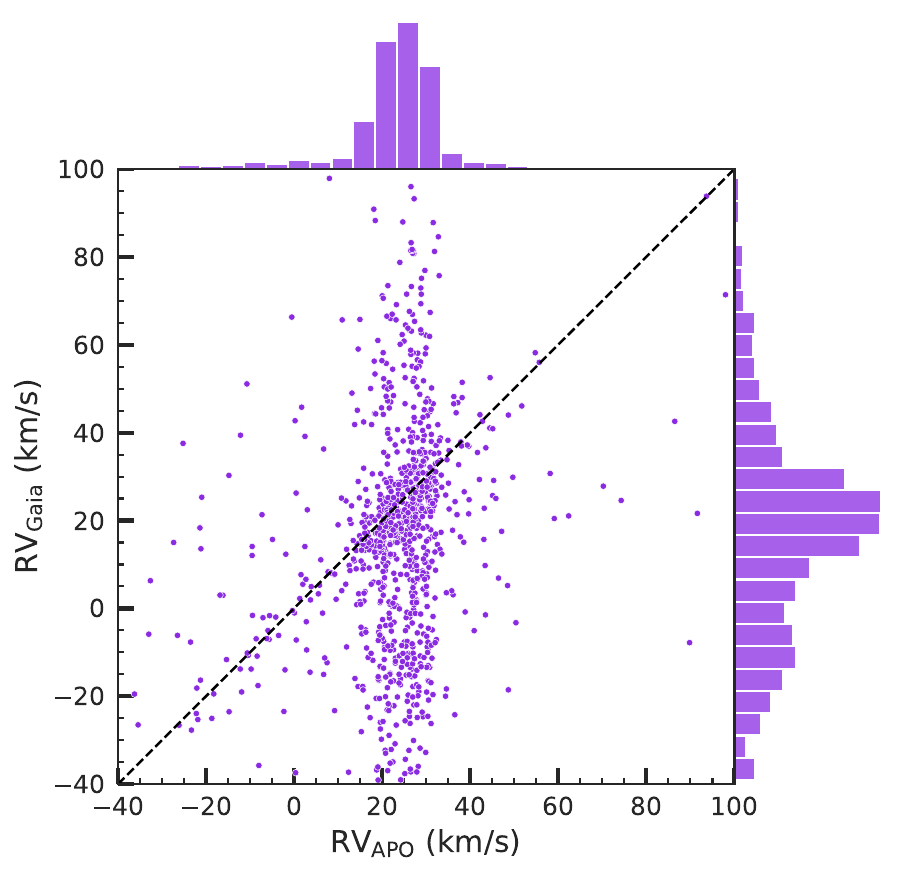}
\caption{Comparison of radial velocities obtained from Gaia DR3 (${\mathrm{RV}}_{\mathrm{Gaia}}$) and radial velocities obtained from APOGEE-2 (${\mathrm{RV}}_{\mathrm{APO}}$). The black dashed line is the one-to-one line.}
\label{appendix_B}
\end{figure}

In the YSO sample mentioned in section \ref{YSOs_sample}, there are 1046 YSO members with available RV information from both Gaia DR3 and APOGEE datasets. In Figure \ref{appendix_B}, we compare their RV measurements from Gaia and APOGEE, where each blue violet point represents a YSO member. We notice a large fraction of YSOs exhibiting conflicting radial velocities between these two datasets. The Pearson correlation coefficient ($r$) of the RV values between them is only 0.26, indicating that the correlation is weak for the sources. Besides, as depicted in Figure \ref{appendix_B}, RVs from Gaia DR3 exhibit a larger dispersion compared to those from APOGEE. These two databases exhibit notable differences, emphasizing the need to prioritize the more accurate one for the RVs of young stars. According to \cite{Kounkel+2023ApJS}, RV measurements for YSOs obtained from Gaia are less accurate when compared to those from APOGEE. One contributing factor is that RV measurements from Gaia rely on accurate solutions designed for ``typical stars", which do not encompass YSOs due to interference from their CaII triplet emission lines in Gaia's spectral range \citep{Recio-Blanco+2023A&A...674A..29R}. These emission lines result from both accretion and activity processes. So we only include the RV measurements derived from APOGEE in the paper.




\end{document}